\documentclass[draft
  ]
  {aipproc}

\layoutstyle{8x11single}

\begin{document}

\title{Rebound Shock Breakouts of Exploding\\
 Massive Stars: A MHD Void Model}


\author{Ren-Yu Hu}{
    address={Physics Department and Tsinghua Center for Astrophysics
    (THCA), Tsinghua University, Beijing 100084, China},
    email={hu-ry07@mails.tsinghua.edu.cn}
}

\author{Yu-Qing Lou}{
  address={Physics Department and Tsinghua Center for Astrophysics (THCA),
  Tsinghua University, Beijing 100084, China},
  email={louyq@mail.tsinghua.edu.cn}
}

\begin{abstract}
With a self-similar magnetohydrodynamic (MHD) model of an
exploding progenitor star and an outgoing rebound shock and with
the thermal bremsstrahlung as the major radiation mechanism in
X-ray bands, we reproduce the early X-ray light curve observed for
the recent event of XRO 080109/SN 2008D association. The X-ray
light curve consists of a fast rise, as the shock travels into the
``visible layer" in the stellar envelope, and a subsequent
power-law decay, as the plasma cools in a self-similar evolution.
The observed spectral softening is naturally expected in our
rebound MHD shock scenario. We propose to attribute the
``non-thermal spectrum" observed to be a superposition of
different thermal spectra produced at different layers of the
stellar envelope.
\end{abstract}

\maketitle


\section{Introduction}

SN 2008D, the best type Ibc supernova detected so far, is preceded
by a X-Ray Outburst (XRO) captured by SWIFT satellite on 2008
January 9, and this XRO is interpreted as a shock breakout of a
Wolf-Rayet (WR) progenitor with a radius of $\sim 10^{11}$ cm
\citep
{Soderberg2008}. The isotropic X-ray
energy is estimated to be $\sim 2\times10^{46}$ erg, and there
seems no collimation detected
so the event is not regarded as a GRB. This XRO showed a rapid
rise, peaked at $\sim 63$ s, and a decay modelled to be
exponential with an e-folding time of $\sim 129$ s
\cite{Soderberg2008}.
The follow-up optical and ultraviolet observations indicate a
total supernova kinetic energy of $\sim 2-4\times10^{51}$ erg and
a mass of SN ejecta to be $\sim 3-5$ M$_{\odot}$
\cite{Soderberg2008}.
Some authors estimate from a
detailed spectral analysis that SN 2008D, originally a $\sim 30$
M$_{\odot}$ star, has a spherical symmetric explosion energy of
$\sim 6\times10^{51}$ erg and an ejected mass $\sim 7$ M$_{\odot}$
\cite{Mazzali}. The evolution of optical spectra of XRO-SN 2008D
resembles that of XRO-SN 2006aj, whose progenitor is also believed
to be a WR star \cite{Campana2006}.

The production of $\gamma$-rays and X-rays by shock breakouts has
been proposed earlier \citep{Colgate, Chevalier}. This XRO and the
associated SN present an unprecedented case to be investigated in
details, especially on interpretations for the rise and decay
times of the X-ray light curve. The claim of an exponential decay
may be premature given a fairly large scatter, and it may have
concealed valuable physical clues offered by this XRO. During the
XRO, the observed spectroscopic softening still lacks a convincing
explanation. Here, we advance a self-similar MHD rebound shock
model in an attempt to reproduce the observed X-ray light curve.
The next section contains an overall description of the
self-similar MHD model and the procedure of analysis; in the third
section, we compare our model results with data; and conclusions
are summed up in the last section.

\section{A Self-Similar MHD Void Shock Model}

For a polytropic magnetofluid in quasi-spherical symmetry under
the self-gravity, the governing magnetohydrodynamic (MHD)
equations include mass conservation, momentum conservation (Euler
equation), magnetic induction equation, and an equation of
specific entropy conservations along streamlines to approximate
energetic processes. For this more general polytropic equation of
state,
we regard the polytropic index $\gamma$ as a parameter
\citep{WangLou08}.

These coupled nonlinear MHD partial differential equations (PDEs)
can be reduced to nonlinear ordinary differential equations (ODEs)
by introducing a self-similar transformation $r=k^{1/2}xt^n$,
where $r$ is the radius, $t$ is the time and $k$ is a scale
parameter relevant to the local sound speed, rendering the
independent self-similar variable $x$ dimensionless. The
corresponding transformation of the dependent MHD variables can be
found in refs. \cite{WangLou08, LouHu}. The exponent $n$ is a key
parameter that determines the dynamic behaviour of a polytropic
fluid. For $n+\gamma=2$, the formulation reduces to that of a
conventional polytropic gas in which the specific entropy remains
constant everywhere \cite{SutoSilk1988,Yahil1983,LouWang06,
LouWang07, WangLou07}.
The special case of $n=1$ and $\gamma=1$ corresponds to the
isothermal case \cite{BianLou2005, YuLouBianWu06}. Such
self-similar evolutions represent an important subclass of all
possible evolutions. We also introduce a dimensionless magnetic
parameter to represent the strength of a magnetic field
$h\equiv<B_t^2>/(16\pi^2G\rho^2r^2)$, where $<B_t^2>$ is the
ensemble average of a random transverse magnetic field squared,
$G$ is the gravity constant and $\rho$ is the mass density.
Meanwhile, MHD shocks are necessary to connect different branches
of self-similar solutions. The conservation laws impose
constraints on physical variables across a MHD shock front. We can
then derive downstream physical quantities (density, velocity,
pressure and temperature) from the upstream physical quantities or
vice versa.
Self-similar solutions produce radial profiles of density, radial
velocity, pressure and temperature at any time of evolution, and
the detailed procedure of analysis can be found in the reference
of Wang \& Lou \cite{WangLou08}. It is also sensible to invoke the
plasma cooling function and obtain radiation diagnostics from a
magnetofluid of high temperatures $\sim 10^7-10^8$K
\cite{Sutherland1993}.

Recently, we obtained a new class of self-similar ``void"
solutions within a certain radius $r^*$ referred to as the void
boundary. In general, such a void solution describes an expanding
fluid envelope with a central cavity and possibly associated with
an outgoing shock \cite{LouCao}. The self-similar evolution
implies that the central void expands as a power-law in time
$r^*\propto t^n$. We study detailed behaviours of void solutions
under different parameters in a general polytropic MHD framework
\citep
{HL, LouHu}. Here, we propose to utilize
such void shock solutions to model the explosion of a massive
progenitor star in the process of a rebound MHD shock breakout.
The Bondi-Parker radius of a remnant compact object if any left in
the center is defined as
\begin{equation}
r_{\rm BP}=\frac{GM_*}{2a^2}\ ,\label{BP}
\end{equation}
where $M_*$ is the mass of the central object and $a$ is the sound
speed at the inner void edge of the surrounding gas. Far beyond
this radius $r_{\rm BP}$, the gravity of the central object
becomes negligible compared to the thermal pressure. For
supernovae, $M_*$ would be of the order of M$_{\odot}$
\cite{Mazzali}. At $\sim 1$ s after the core bounce, the
temperature of the stellar envelope is of the order of $10^8$ K,
and the sound speed $a^2\sim10^{17}$ cm$^2$ s$^{-2}$, and then
$r_{\rm BP}\sim 10^{8}$ cm. Meanwhile,
the void radius $r^*$ expands to larger than $10^8$ cm
\citep{Janka}. Furthermore, the Bondi-Parker radius expands slower
than the void boundary does \cite{LouHu}.
Therefore, the cavity assumption may be justifiable.

\section{MHD Model and X-ray Light Curve}

\begin{figure}
  \includegraphics[height=.3\textheight]{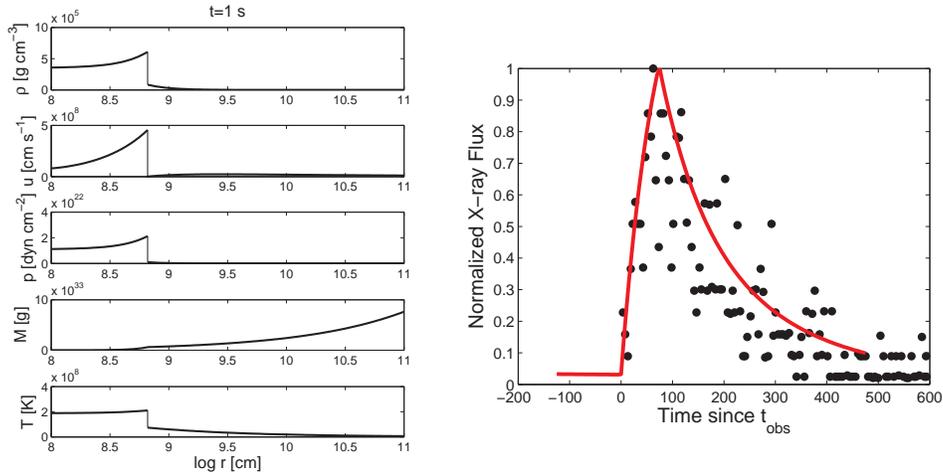}
  \caption{Our MHD void shock model for a shock breakout in a
  progenitor of SN 2008D (left) and the resulting X-ray light
  curve (right). On the left from top to bottom, the panels
  show the radial profiles of density, radial velocity, pressure,
  enclosed mass and temperature of the stellar envelope within
  radial range $10^8$ cm (void boundary) and $\sim10^{11}$ cm
  (outer boundary) at 1 s after the core collapse and rebounce.
  The model is obtained with the self-similar parameters as $n=0.8$,
  $\gamma=1.2$ (conventional polytropic) and $h=0$ (non-magnetized
  fluid).
  On the right, we compare the X-ray light curve calculated from our MHD
  void shock model (red curve) and data from the X-Ray Telescope (XRT)
  on board the SWIFT satellite \cite{Soderberg2008} (solid circles with
  error bars suppressed). X-ray fluxes are normalized to the peak flux.
  The X-ray light curve is shown as a function of time since the XRT
  trigger, noted as $t_{\rm obs}$. The core collapse happened
  $\sim 552$ s before the trigger.
  The X-ray light curve
  is calculated with the radiation layer thickness to be $9\times10^{9}$ cm. The
  calculated X-ray light curve is in a shape of fast-rise-and-decay. The rise time
  is $\sim 62$ s, and the decay obeys a power law to the time since core collapse
  with the index to be $-4.3$. The equivalent e-folding is 128 s (i.e., the
  timescale when the emission intensity drops to $\sim 37\%$ of the peak value).}
  \label{Figure1}
\end{figure}

The self-similar MHD void shock model of a WR stellar envelope in
explosion associated with a shock breakout and the corresponding
X-ray light curve are shown in Figure \ref{Figure1}.

Following observational inferences \cite{Soderberg2008} for the
progenitor radius, we cut off our model at this ``outer boundary"
($r_{\rm out}\sim 10^{11}$ cm). This approximation is reasonable,
as the radial density profile of the star drops rapidly at the
stellar surface. We do not consider dynamical effects and the
X-ray contributions of the gas outside $r_{\rm out}$. Our MHD
model gives an enclosed mass at $\sim 10^{11}$ cm to be $3.8$
M$_{\odot}$, comparable to the estimated mass of ejecta $(\sim
3-5$ M$_{\odot})$. The gas kinetic energy is $\sim 3\times10^{51}$
erg, also in the observed range. The gravitational binding energy
given by our model is $\sim 10^{50}$ erg, much less than the
kinetic energy corresponding to an exploding stellar envelope
during a MHD rebound shock breakout.

In Figure \ref{Figure1}, we see features of a rebound MHD shock
surrounding a central void in self-similar expansion. From the
upstream to downstream sides across the shock, the density,
pressure and temperature increases suddenly. The radial velocity
also increases, but the velocity in the shock comoving reference
framework decreases as expected from the upstream to downstream
sides. Note that the temperature is $\sim 10^8$ K and it drops to
$\sim 10^7$ K in the process of the evolution under consideration,
corresponding to the energy range of X-ray photons detected.
Typically the part of gas near the downstream shock front, where
the density and temperature are the highest, is most efficient in
producing X-ray emissions. We suggest that X-ray emissions
observed are from the thermal bremsstrahlung radiation mainly
produced around the downstream side of a rebound shock.

We compute the X-ray light curve using the plasma cooling rate
result of reference \cite[][see also Lou \& Zhai 2008 in
preparation for X-ray diagnostics of isothermal voids in
self-similar expansion]{Sutherland1993}, in which both the
free-free and free-bound emissions are taken into account. The
optical depth in the stellar envelope is unknown, so we treat it
as another parameter to search for the best fit of X-ray light
curve. Here we introduce an ``inner boundary" $r_{\rm in}$, and
presume that only X-ray emissions in the layer between $r_{\rm
in}$ and $r_{\rm out}$ can be observed and should be integrated.
The thickness of such ``radiative layer" noted as $s$ is another
parameter to adjust.

Our scenario is as follows. The MHD rebound shock front expands
outward obeying a power law in time $t$ since the core collapse,
and the shock strength weakens with increasing $t$. Before the
shock front reaches $r_{\rm in}$, the density and temperature are
low in the radiative layer and cannot produce detectable X-ray
emissions. Once the shock front reaches $r_{\rm in}$ and runs into
the radiative layer, more and more downstream part enters the
radiative layer, and X-ray emissions increase rapidly. The X-ray
emission reaches its maximum when the shock front reaches $r_{\rm
out}$ (shock breakout) and the entire radiative layer is occupied
by the downstream part. Thereafter, the density and temperature
inside the radiative layer decrease self-similarly, and the X-ray
emission decreases obeying a power law. The radiative layer
thickness $s$ and the shock speed determine the rise time and the
power law index of the subsequent decay. The average temperature
of the radiative layer decreases self-similarly in the shock
breakout process, naturally leading to the spectral softening as
observed.

\section{Summary and Conclusions}

With a self-similar MHD void shock model and the thermal
bremsstrahlung as the main radiation loss, we obtain a fairly good
fit to the X-ray light curve observed and confirm that XRO 080109
is most likely a shock breakout event. We identify that the decay
in the X-ray light curve follows a power law (instead of an
exponential law) in time since the core collapse and rebounce,
which occurred $\sim 552$ s before the observation of X-ray
emissions. Meanwhile, the spectral softening is expected
qualitatively.
In this work, we use the most simplified radiation transfer
presumption that the radiation produced in the `visible layer' can
be totally observed. Actually, the optical depth varies with
radius and should be treated in a more elaborate manner.
Additionally, we presume that the boundaries of the ``visible
layer" $r_{in}$ and $r_{out}$ do not vary with time. Despite all
these idealizations, our self-similar dynamic approach appears to
be suitable to couple with the radiation process and to model
X-ray outbursts in supernova as observed.

Regarding the X-ray spectra observed, they cannot be fitted with a
simple blackbody profile,
and a nonthermal power law profile was suggested
\cite{Soderberg2008}.
Several mechanisms have been
proposed to explain such X-ray spectra, for example the bulk
comptonization by scatterings of the photons between the ejecta
and a dense circumstellar medium \cite{Soderberg2008}, or diluted
thermal spectra which require the thermalization occurs at a
considerable depth in the supernova \cite{Chevalier2008}. We
propose that the power-law profile might be a natural result of
multi-colour superposition of blackbody spectra.
Based on our
scenario outlined here, X-ray emissions come from different layers
within the radiative layer around the stellar surface, and the
radiative layer has different temperatures at different depth. As
a result, the observed X-ray spectra are the superposition of
thermal blackbody components with different temperatures. We
suggest that this might resolve issues of spectral profile and
evolution.

During breakouts of rebound shocks and in the presence of MHD
shock accelerated relativistic electrons usually presumed with a
power-law energy spectrum within a certain electron energy range,
we could also compute synchrotron emissions associated with such
kind of SN shock breakouts. Among others, it is then possible to
follow the evolution of magnetic field strength associated with SN
explosions \cite{Lou94, LouWang07} and estimate the effectiveness
of accelerating relativistic particles (i.e., high-energy cosmic
rays \cite{Science06}). There is the freedom of choosing a few
parameters to fit the data at a certain epoch. It is then possible
to test the hypothesis of a self-similar shock evolution by
further observations. In a more general perspective and on the
basis of our dynamic models for rebound MHD shocks, we hope to
further develop radiative diagnostics for shock breakouts of
supernovae and thus for SN related GRBs.


\begin{theacknowledgments}
This research has been partially supported by Tsinghua Center for
Astrophysics (THCA), by the National Natural Science Foundation of
China (NSFC) grants 10373009 and 10533020 and by the National
Basic Science Talent Training Foundation
(NSFC J0630317) at the Tsinghua University, and by the SRFDP
20050003088 and the Yangtze Endowment from the Ministry of
Education at Tsinghua University.
\end{theacknowledgments}

\end{document}